\documentclass[letterpaper,10pt,conference]{ieeeconf}
\IEEEoverridecommandlockouts  %
\usepackage{amsmath,amssymb,amsfonts}

\usepackage{amsthm}
\usepackage{graphicx}
\usepackage{booktabs}
\usepackage[caption=false,font=footnotesize]{subfig} %
\usepackage{xcolor}
\newcommand{\captabgap}{\vspace{-5pt}} %
\usepackage{cite}
\usepackage[colorlinks=true,allcolors=blue]{hyperref}

\newtheorem{lemma}{Lemma}
\newtheorem{theorem}{Theorem}
\newtheorem{proposition}{Proposition}
\newtheorem{corollary}{Corollary}
\newtheorem{assumption}{Assumption}
\theoremstyle{remark}
\newtheorem{remark}{Remark}

\DeclareMathOperator{\col}{col}
\newcommand{\Rset}{\mathbb{R}}
\newcommand{\Tini}{T_{\mathrm{ini}}}
\newcommand{\yini}{\bar y_{\mathrm{ini}}}
\newcommand{\uini}{\bar u_{\mathrm{ini}}}
\newcommand{\dini}{d_{\mathrm{ini}}}
\newcommand{\yc}[1]{\hat{\bar y}^{\,c}_{t+#1}}
\newcommand{\err}{\rho}
\newcommand{\trans}{^{\!\top}}
\newcommand{\band}{\bar\omega}
\newcommand{\SC}{SC-DeePC}
\makeatletter
\@ifundefined{IEEEkeywords}{\newenvironment{IEEEkeywords}%
  {\par\noindent\textbf{\textit{Keywords}---}}{\par}}{}
\makeatother
\definecolor{blue}{rgb}{0,0,0}

\usepackage{eso-pic}
\AddToShipoutPictureBG*{\AtPageUpperLeft{\put(\LenToUnit{0.5\paperwidth},\LenToUnit{-0.42in}){\makebox[0pt][c]{\parbox{0.82\paperwidth}{\centering\footnotesize This work has been submitted to the IEEE for possible publication. Copyright may be transferred without notice, after which this version may no longer be accessible.}}}}}
\begin{document}

\title{Koopman-Corrected DeePC with Conformal Safety\\ Certificates for Transient Frequency Control}

\author{Amin Taghieh and SangWoo Park%
\thanks{The authors are with New Jersey Institute of Technology, Newark,
NJ 07102 USA (e-mail: at2228@njit.edu; sangwoo.park@njit.edu).}}

\maketitle
\thispagestyle{empty}
\pagestyle{empty}

\begin{abstract}
Data-enabled predictive control (DeePC) represents future trajectories
through recorded input--output data. Although this representation is
exact for noise-free linear systems under suitable conditions,
prediction errors arise when DeePC is applied to nonlinear plants,
potentially invalidating output constraints. This paper develops
safety-certified DeePC (\SC), which constrains a corrected data-driven
prediction using learned uncertainty margins. A nominal Koopman
model of the plant, identified by extended dynamic mode decomposition
(EDMD) on data of the intact network, yields an exact decomposition of the corrected prediction
error into the error of estimating the current lifted state from the
measured window and the model residuals along the horizon.
Deterministic worst-case bounds are derived, but they are guaranteed
only while the model residuals stay within bounds measured on the
intact network, which large disturbances can violate. Borrowing the
concept of conformal robustness, a recurrent neural network instead
predicts a window-dependent error scale, and trajectory-level split
conformal calibration makes the resulting margin valid simultaneously
over a whole scenario under scenario-level exchangeability. Tightening
the output constraint on the corrected prediction by this margin yields
a conditional one-step safety certificate at
instants where the tightening is active and no safety relaxation is
required. The method is evaluated on transient frequency control of the
IEEE nine-bus power system.
\end{abstract}

\begin{IEEEkeywords}
Data-driven predictive control, nonlinear systems, conformal
prediction, Koopman operator, power system frequency control.
\end{IEEEkeywords}

\section{Introduction}
Data-enabled predictive control (DeePC)
\cite{coulson2019deepc} replaces the model of a predictive controller
by a record of input--output data. By the fundamental lemma
\cite{willems2005note}, a single persistently exciting record of a
linear system parameterizes all of its trajectories of a given length.
Regularization and distributional robustness were added to DeePC in
\cite{coulson2020regularization}, and closed-loop stability and
robustness of a data-driven predictive controller were proved in
\cite{berberich2021data}. For nonlinear systems, Koopman operator theory
offers a linear representation in a lifted space, approximated in
practice by extended dynamic mode decomposition (EDMD)
\cite{kordamezic2018}, but the choice of lifting functions is difficult
and can introduce a bias. The two views are bridged in
\cite{koopmanlemma2024}, which shows that, for nonlinear systems that
admit a Koopman linear embedding, the trajectory space of the embedding
is spanned by rich-enough trajectories of the nonlinear system, so that
a data-driven representation can bypass the lifting functions. It also
shows that both the width and the depth of the trajectory library
matter. Practical
exponential stability of data-driven predictive control applied directly
to input--output data of a nonlinear system was established in
\cite{taghieh2026lcss}, through an approximate Koopman embedding that
serves only as a theoretical certificate. The model-based route is
followed in \cite{strasser2026safedmd}, with an EDMD surrogate that has
proportional error bounds and a controller with guaranteed
stabilization. Conformal prediction has
recently given distribution-free bounds on the modeling error along
closed-loop trajectories, for data-driven Lyapunov and barrier
certificates \cite{hsu2025conformal}, for Koopman-based control
\cite{hirano2026conformalkoopman}, and for safe planning among
dynamic agents \cite{lindemann2023safe}.

These developments leave open how to obtain run-time certificates of
output-constraint satisfaction for DeePC applied to nonlinear plants
with state-dependent prediction errors. The usual route in robust predictive control is to tighten
the constraint on a prediction by a bound on its error. The bound must
be valid and small compared with the admissible band, since otherwise no
input can meet the tightened constraint. A bound valid for every
operating condition of a nonlinear plant, however, is often far wider.

The application is transient frequency control of a multi-machine power
network. DeePC has been used for power-system oscillation damping
\cite{huang2021decentralized}, and Koopman MPC for transient
stabilization \cite{korda2018powergrid}. After a fault, the frequency of every generator must stay
inside a band, and after a line outage the network is no longer the
system on which the data were recorded, so a DeePC prediction built from
trajectories recorded before the outage has no reason to predict the
post-outage plant. This paper takes the structure of the prediction
error from a model and its size from data. A nominal Koopman
model of the plant, the EDMD approximation of the Koopman operator on a
chosen set of observables (the nominal system of
\cite{taghieh2026lcss}), identified on data of the intact network,
yields a corrected
multi-step prediction of the subspace-predictive-control form
\cite{fiedler2021relationship}. At every instant and prediction step,
its error on the nonlinear plant equals the model residuals accumulated
since the start of the measured window minus a linear map of their
values over the window. To size this error, we borrow the concept of
conformal robustness from \cite{hsu2025conformal}, where Lyapunov and
barrier certificates are made robust to model uncertainty
quantified by trajectory-level conformal prediction, and carry it to
DeePC through the exact error identity. A recurrent network learns a
window-dependent error scale, in the spirit of conformal time-series
forecasting \cite{stankeviciute2021conformal}, and trajectory-level
split conformal prediction
\cite{vovk2005algorithmic,angelopoulos2023conformal}, with one score per
closed-loop scenario, calibrates this scale into a margin with
simultaneous coverage over a new exchangeable scenario. The controller
that tightens its output constraint by this margin is called
safety-certified DeePC (\SC). Certification is issued only when the
learned tightening is active and met without relaxation, and its
probability rests on the calibration assumptions.

\begin{figure*}[t]
\centering
\includegraphics[width=0.9\textwidth]{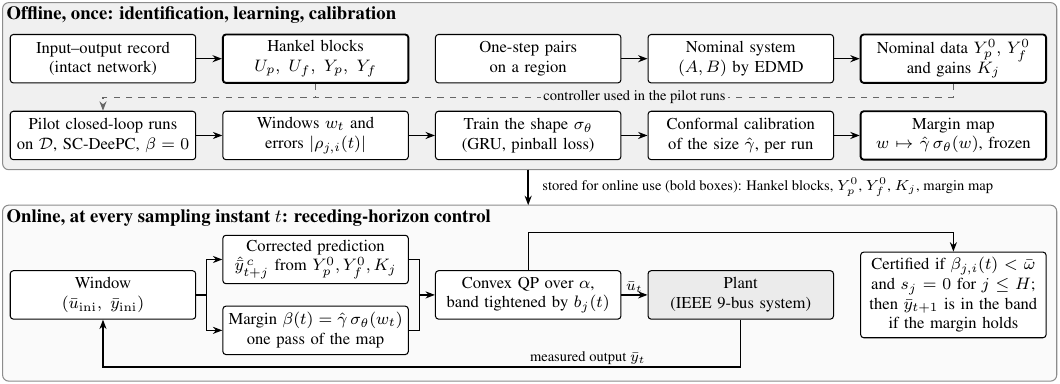}
\caption{Method overview. Offline, EDMD fits the nominal Koopman
model, the
input--output record gives the Hankel blocks, and closed-loop scenarios
of \SC\ with $\beta=0$ give the errors on which the shape
$\sigma_\theta$ is trained and the size $\hat\gamma$ is calibrated by
conformal prediction.
Online, each instant needs one pass of the margin network and one
convex QP. A certified
instant is followed by an in-band sample whenever the margin holds,
which Theorem~\ref{thm:cp} guarantees with probability at least
$1-\epsilon$ under Assumption~\ref{as:exch} (Remark~\ref{rem:shift}).}
\label{fig:overview}
\end{figure*}

\paragraph*{Contributions}
First, an exact expression for the error of a corrected DeePC
prediction on a nonlinear plant (Lemma~\ref{lem:error}), its
split into a window part and a horizon part
(Corollary~\ref{cor:struct}) and the exact worst case over a box of
model residuals (Proposition~\ref{prop:wc}), which identify what a
margin must cover and why it should depend on the measured window.
Second, conformal robustness \cite{hsu2025conformal} extended to DeePC
through this identity: a learned, window-dependent margin on the
corrected prediction error, calibrated with the closed-loop scenario as
the unit, so that the standard split conformal guarantee holds
simultaneously over every instant of a new scenario under
scenario-level exchangeability (Theorem~\ref{thm:cp}).
Third, \SC, which keeps the structure of
regularized DeePC and certifies the next output sample whenever the
tightened constraint is met without relaxation
(Proposition~\ref{prop:cert}).
Online, the controller uses only precomputed matrices, the margin
network and the measured window, and nothing in the method uses the
admittance matrix, the inertias or the internal voltages; offline, the
nominal Koopman model is fitted on state-space samples from a
simulator (Fig.~\ref{fig:overview}).
In contrast with \cite{taghieh2026lcss}, where a Koopman embedding
serves only as a theoretical certificate, the nominal Koopman model is
used here constructively: it defines the corrected prediction and the
exact identity for its error. Removing the correction and the
margin returns regularized DeePC (Remark~\ref{rem:why}).

\section{Problem Formulation}
\label{sec:plant}
\vspace{-1mm}
\subsection{Plant and safety requirement}

The plant considered in this study is the IEEE 9-bus system, which has three synchronous generators~\cite{anderson2003power}. For each machine $i\in\{1,2,3\}$, the dynamics can be expressed by the following equations
\vspace{-1mm}
\begin{align}
  \dot\delta_i &= \Omega_b(\omega_i-\omega_s),\label{eq:swing1}\\
  M_i\dot\omega_i &= P_{m0,i}+u_i-P_{e,i}(\delta),
  \label{eq:swing2}\\
  P_{e,i}(\delta) &= E_i^2G_{ii}+\textstyle\sum_{j\neq i}E_iE_j
  \bigl(G_{ij}\cos\delta_{ij}+B_{ij}\sin\delta_{ij}\bigr),\label{eq:pe}
\end{align}

\vspace{-2mm}
\noindent with $\delta_{ij}=\delta_i-\delta_j$, $u_i$ a controllable power
injection, and $G+\mathrm{j}B$ the Kron-reduced admittance matrix. The
only nonlinearity is \eqref{eq:pe}. There is no damping term, so all
damping comes from the controller.

Deviations from the operating point are denoted by
$\bar u=u-u_s$ and $\bar y=y-y_s$, with the measured output
$\bar y=\col(\omega-\omega_s,\ \delta-\delta_0)\in\Rset^{6}$. Let $S$
select its three frequency rows. The safety requirement is
\begin{equation}
  |(S\bar y_k)_i|\le\band\quad\text{for all machines $i$ and all $k$},
  \label{eq:safety}
\end{equation}
with $\band$ the frequency band. Inputs are limited in magnitude,
$|u_i|\le u_{\max}$, and in rate, $|\Delta u_i|\le\Delta u_{\max}$ per
sample.

The events that call for transient frequency control include
short-circuit faults and transmission-line outages, as well as sudden
load changes. A three-phase fault
at a bus is a low-impedance path from that bus to ground, modeled by a
shunt admittance $y_f$ added to the corresponding diagonal entry of the
bus admittance matrix before Kron reduction. Protection clears the fault after a
short time, either by removing the shunt or by opening the breakers of
the faulted line, which causes a line outage, i.e., removes its
branch admittance. Each event changes $G+\mathrm{j}B$ in \eqref{eq:pe}.
During a fault, the generators deliver less electrical power and
accelerate, so their frequencies leave the band unless the inputs act;
after an outage, the network differs permanently from the one on which
the data were recorded. Our goal is a DeePC controller
that keeps \eqref{eq:safety} during and after such events and, from
measured data alone, certifies at run time, with a prescribed
probability, that the next frequency sample satisfies it.

\subsection{Sampling instants and windows}
\label{sec:timing}

The controller operates at discrete \emph{sampling instants}
$t=0,1,2,\dots$, one sampling period apart. At instant $t$ the output
$\bar y_t$ has just been measured and the input $\bar u_t$ is about to
be chosen; $\bar u_t$ acts during the following period, so the first
output it influences is $\bar y_{t+1}$. The controller uses the
\emph{window} of the last $\Tini$ complete input--output pairs,
\begin{equation}
\begin{aligned}
  \uini&=\col(\bar u_{t-\Tini},\dots,\bar u_{t-1}),\\
  \yini&=\col(\bar y_{t-\Tini},\dots,\bar y_{t-1}),
\end{aligned}
  \label{eq:window}
\end{equation}
each output being the one measured before its input acted; the recorded
data are paired the same way. Predictions are indexed by the number of
steps ahead, $\hat{\bar y}_{t+j}$ for $j=0,\dots,N-1$; step $j=0$ predicts
the current measurement, which $\bar u_t$ cannot change.

\subsection{Regularized DeePC and problem statement}

From a recorded input--output sequence of the intact network, build
block Hankel matrices of depth $\Tini+N$ and split them into past and
future rows, $U_p,U_f,Y_p,Y_f$. Regularized DeePC
\cite{coulson2019deepc,coulson2020regularization} finds a combination
$\alpha$ of the recorded columns such that
\begin{equation}
  U_p\alpha=\uini,\qquad Y_p\alpha=\yini+\sigma_p,\qquad
  U_f\alpha=\bar u_f,
  \label{eq:consist}
\end{equation}
and predicts $\hat{\bar y}_{t+j}=E_jY_f\alpha$, where $E_j$ selects
block $j$ and $\bar u_f=\col(\bar u_t,\dots,\bar u_{t+N-1})$ is the
planned input. The slack $\sigma_p$ and a penalty on $\|\alpha\|$ make
the problem well posed under noise and nonlinearity, heuristically. The
prediction is exact when the plant is linear and the record is
persistently exciting; on \eqref{eq:swing1}--\eqref{eq:pe} it is not.
We call this baseline DeePC. The problem is to design a controller on
the same data that, at every sampling instant, either certifies that the
next frequency sample satisfies \eqref{eq:safety} or declares that it
cannot, with certificates that hold with a prescribed probability and
are issued often enough to be useful.

\section{A Corrected Prediction With an Exact Error}
\label{sec:pred}

This section builds the prediction that \SC\ constrains and
characterizes its error exactly. The plant is written as a linear model
in a lifted space, the \emph{nominal Koopman model}, plus a
residual that collects everything this model misses. The nominal
Koopman model enters the controller only through matrices
precomputed offline. Lifting functions are evaluated only offline, to fit
the model and to start each Hankel column of the nominal model at its
lifted state. Online, the current lifted state is not computed from the
observables but estimated implicitly from the measured input--output
window, so only the rotor angles and speeds need to be measured.
Section~\ref{sec:corr} corrects the DeePC prediction so that, for a
fixed input, its error depends on neither $\alpha$ nor $\sigma_p$, and
Lemma~\ref{lem:error} expresses this error exactly in the residuals.
Section~\ref{sec:struct} shows that deterministic bounds on the error
are guaranteed only while the residuals stay within bounds measured on
the intact network, and that the error splits into a part fixed by the residuals
inside the window and a part from the residuals along the horizon,
which motivates a margin that reads the window
(Section~\ref{sec:bound}).

\vspace{-1mm}
\subsection{\texorpdfstring{Nominal Koopman model}{Nominal Koopman model} and accumulated residual}

Let $\Phi$ be a vector of $n_z$ observables of the state and
$\bar z_k=\Phi(x_k)-\Phi(x_s)$ the lifted state. If $\operatorname{span}\Phi$
were invariant under the Koopman operator of
\eqref{eq:swing1}--\eqref{eq:pe}, the lifted state would evolve
linearly under the input. This fails for the swing dynamics, so
we use the EDMD approximation of the Koopman operator on
$\operatorname{span}\Phi$ \cite{kordamezic2018}, called the \emph{nominal
Koopman model} (the nominal system of \cite{taghieh2026lcss}),
\vspace{-2mm}
\begin{equation}
  \bar z^0_{k+1}=A\bar z^0_k+B\bar u_k,\qquad \bar y^0_k=C\bar z^0_k ,
  \label{eq:N}
\end{equation}

\vspace{-1mm}
\noindent whose pair $(A,B)$ is fitted by least squares \cite{kordamezic2018} on
\emph{one-step pairs}, i.e., samples $(x_k,u_k,x_{k+1})$ in which
$x_{k+1}$ is the state reached from $x_k$ after one sampling period
under the constant input $u_k$, drawn on a region around the operating
point of the intact network rather than along trajectories. The output
is part of $\Phi$, so $\bar y_k=C\bar z_k$ with $C$ a selection matrix.
The plant differs from the nominal Koopman model by the residual
\begin{equation}
  e_k:=\bar z_{k+1}-A\bar z_k-B\bar u_k ,
  \label{eq:resid}
\end{equation}
which collects three effects: the failure of $\operatorname{span}\Phi$ to be
Koopman-invariant, the EDMD fitting error, and the effect of
any topology change caused by a fault or a line outage, since $(A,B)$
is fitted on the intact network.

Start the nominal model at the true lifted state at the start of
the window, $\bar z_{t-\Tini}$, and drive it with the plant's inputs. By \eqref{eq:N} and \eqref{eq:resid}, the plant output $j$
steps later differs from the nominal output by the \emph{accumulated
residual}
\vspace{-1mm}
\begin{equation}
  d_j:=\sum_{s=0}^{j-1}CA^{\,j-1-s}e_{t-\Tini+s},\qquad d_0=0 .
  \label{eq:drift}
\end{equation}
Let $\dini=\col(d_0,\dots,d_{\Tini-1})$ collect it over the window. It
is unknown, but it has already occurred and is contained in the
measured window $\yini$.

\subsection{Corrected prediction and the gains \texorpdfstring{$K_j$}{Kj}}
\label{sec:corr}

The corrected prediction is obtained in three steps. First, let
$Y^0_p,Y^0_f$ be Hankel matrices of the nominal model's output on the
recorded inputs, each column started from the measured lifted state at
its first sample, so that every column of
$\col(U_p,U_f,Y^0_p,Y^0_f)$ is an exact trajectory of \eqref{eq:N}.
Since \eqref{eq:N} is linear, every combination of columns is one
too, so for each $\alpha$ there is an initial lifted state
$\zeta_\alpha$ with
\begin{equation}
  Y^0_p\alpha=\mathcal{O}\zeta_\alpha+TU_p\alpha,\quad
  E_jY^0_f\alpha=CA^{\Tini+j}\zeta_\alpha+G_j\mathcal{U}\alpha ,
  \label{eq:nomtraj}
\end{equation}
where $\mathcal{O}=\col(C,CA,\dots,CA^{\Tini-1})$ is the observability
matrix, $\mathcal{U}:=\col(U_p,U_f)$, and $T$ and $G_j$ collect the
Markov parameters $CA^qB$ that map the inputs to the window and to step
$j$. Second, by \eqref{eq:resid} and \eqref{eq:drift}, the plant
satisfies the same relations plus the accumulated residual. If it
applies $\uini$ and the
planned inputs, with $\bar u:=\col(\uini,\bar u_f)$,
\vspace{-1mm}
\begin{equation}
\begin{aligned}
  \yini&=\mathcal{O}\bar z_{t-\Tini}+T\uini+\dini,\\
  \bar y_{t+j}&=CA^{\Tini+j}\bar z_{t-\Tini}+G_j\bar u+d_{\Tini+j}.
\end{aligned}
\label{eq:plantwin}
\end{equation}

\vspace{-2mm}
\noindent The plant's initial state $\bar z_{t-\Tini}$ is unknown, and
$\zeta_\alpha$ depends on the decision variable $\alpha$. Third, the
gains
\vspace{-1mm}
\begin{equation}
  K_j:=CA^{\Tini+j}\,\mathcal{O}^{\dagger},
  \quad\text{so that}\quad K_j\,\mathcal{O}=CA^{\Tini+j}
  \label{eq:K}
\end{equation}
whenever $\mathcal{O}$ has full column rank, map an input-free output
window to the output it implies at step $j$ of \eqref{eq:N}; among all
such gains, $K_j$ has the smallest norm in every row. Applying $K_j$ to
the first relation in
\eqref{eq:nomtraj} and subtracting it from the second removes
$\zeta_\alpha$ exactly, and adding $K_j\yini$ replaces it by an estimate
of the plant's state from the measured window. This gives the
\emph{corrected prediction}
\begin{equation}
  \yc{j}:=E_jY^0_f\alpha-K_jY^0_p\alpha+K_j\,\yini ,
  \label{eq:corrected}
\end{equation}
which, with $U_p\alpha=\uini$ and
$\bar u_f=U_f\alpha$, equals
\begin{equation}
  \yc{j}=CA^{\Tini+j}\hat z+G_j\bar u,\qquad
  \hat z:=\mathcal{O}^{\dagger}(\yini-T\uini).
  \label{eq:zhat}
\end{equation}
The corrected prediction is thus the output of the nominal model
started from $\hat z$, the least-squares estimate of
$\bar z_{t-\Tini}$ from the measured window, and driven by the past and
planned inputs; the recorded data enter only through the planned input
$U_f\alpha$, and the error of $\hat z$ is where the accumulated
residual over the window enters the prediction error. Denote the prediction error by
\begin{equation}
  \err_j(t):=\bar y_{t+j}-\yc{j},\qquad \err_{j,i}(t)\ \text{its $i$-th
  entry}.
  \label{eq:rho}
\end{equation}

\begin{lemma}[]\label{lem:error}
If $\mathcal{O}$ has full column rank, $U_p\alpha=\uini$ and
$U_f\alpha=\bar u_f$, and the plant applies the
inputs $\uini$ and $\bar u_t,\dots,\bar u_{t+j-1}$, then
\begin{equation}
  \err_j(t)=d_{\Tini+j}-K_j\,\dini .
  \label{eq:error}
\end{equation}
\end{lemma}
\begin{proof}
By \eqref{eq:plantwin} and \eqref{eq:zhat},
$\err_j(t)=CA^{\Tini+j}(\bar z_{t-\Tini}-\hat z)+d_{\Tini+j}$, and the
first relation in \eqref{eq:plantwin} gives
$\hat z-\bar z_{t-\Tini}=\mathcal{O}^{\dagger}\dini$ because
$\mathcal{O}^{\dagger}\mathcal{O}=I$. With \eqref{eq:K} this is
\eqref{eq:error}.
\end{proof}

Three facts follow. First, the error contains neither $\alpha$ nor
$\sigma_p$, so the margin must bound only the accumulated
residual; by \eqref{eq:zhat}, $\yc{j}$ is a
predictor of the subspace-predictive-control type
\cite{fiedler2021relationship} (Remark~\ref{rem:why}). Second,
$K_j\dini$ is fixed by residuals that have already occurred, so the
window carries information about it (Section~\ref{sec:struct}). Third, measurement noise $v$ in the window
enters the error as $-K_jv$, and the observables set the size of
$K_j$ (Section~\ref{sec:obs}).

\subsection{Worst-case bounds and the structure of the error}
\label{sec:struct}

Lemma~\ref{lem:error} can be used in two ways. The deterministic way
bounds every residual. Inserting \eqref{eq:drift} into
\eqref{eq:error} makes the error linear in the residuals
over the window and horizon,
\begin{equation}
  \err_j(t)=\sum_{s=0}^{\Tini+j-1}\Gamma_{j,s}\,e_{t-\Tini+s},
  \label{eq:gamma}
\end{equation}
with
$\Gamma_{j,s}:=CA^{\Tini+j-1-s}-\sum_{q=s+1}^{\Tini-1}K^{(q)}_jCA^{q-1-s}$,
with $K^{(q)}_j$ the $q$-th block column of $K_j$ and an empty sum
zero. Let $\bar e$ bound the residual entrywise over the operating
region; absolute values and inequalities are entrywise.

\begin{proposition}[Worst-case bounds]\label{prop:wc}
Under the conditions of Lemma~\ref{lem:error}, if $|e_k|\le\bar e$ for
$k=t-\Tini,\dots,t+j-1$, then
\begin{align}
  |\err_j(t)|&\le\beta^{\mathrm{ex}}_j
  :=\sum_{s=0}^{\Tini+j-1}|\Gamma_{j,s}|\,\bar e\nonumber\\
  &\le\beta^{\mathrm{sep}}_j:=\bar d(\Tini+j)
  +\sum_{q=0}^{\Tini-1}|K^{(q)}_j|\,\bar d(q),
  \label{eq:det}
\end{align}

\vspace{-2mm}
\noindent where $\bar d(q):=\sum_{r=0}^{q-1}|CA^{r}|\,\bar e$ bounds $|d_q|$. Each
entry of $\beta^{\mathrm{ex}}_j$ is attained by a residual sequence in
the box, so $\beta^{\mathrm{ex}}_j$ is the worst case over the box.
\end{proposition}
\begin{proof}
Inserting \eqref{eq:drift} in \eqref{eq:error} gives \eqref{eq:gamma},
so $|\err_{j,i}(t)|\le\sum_s(|\Gamma_{j,s}|\,|e_{t-\Tini+s}|)_i
\le(\beta^{\mathrm{ex}}_j)_i$, with equality when each residual entry
is $\bar e$ times the sign of the matching entry of row $i$ of
$\Gamma_{j,s}$. The second inequality follows from
$|\Gamma_{j,s}|\le|CA^{\Tini+j-1-s}|+\sum_{q=s+1}^{\Tini-1}|K^{(q)}_j|\,|CA^{q-1-s}|$
by summing over $s$ and exchanging the sums over $s$ and $q$.
\end{proof}

The bound $\beta^{\mathrm{sep}}$ treats the two terms of
\eqref{eq:error} separately and ignores that they partly cancel. Both
bounds ignore the window and are guaranteed only while
$|e_k|\le\bar e$, which holds, up to sampling, on the region of the
intact network where $\bar e$ was measured and fails when a fault or a
line outage drives the residuals out of the box, exactly when a margin
is needed (Section~\ref{sec:results}).
The second way asks where the error comes from at each instant. Let
$\mathcal{R}:=[A^{\Tini-1}B\ \cdots\ AB\ B]$ and
$\hat z_t:=A^{\Tini}\hat z+\mathcal{R}\uini$, the window estimate of
$\bar z_t$.

\begin{corollary}[]\label{cor:struct}
Under the conditions of Lemma~\ref{lem:error},
\begin{equation}
  \err_j(t)=CA^{j}(\bar z_t-\hat z_t)
  +\sum_{q=0}^{j-1}CA^{\,j-1-q}e_{t+q},
  \label{eq:struct}
\end{equation}
where the estimation error
\begin{equation}
  \bar z_t-\hat z_t=\sum_{s=0}^{\Tini-1}A^{\Tini-1-s}e_{t-\Tini+s}
  -A^{\Tini}\mathcal{O}^{\dagger}\dini
  \label{eq:est}
\end{equation}
depends only on the residuals $e_{t-\Tini},\dots,e_{t-1}$. In
particular, $\err_1(t)=CA(\bar z_t-\hat z_t)+Ce_t$.
\end{corollary}
\begin{proof}
Iterating \eqref{eq:resid} from $t$ with the planned inputs gives
$\bar y_{t+j}=CA^j\bar z_t+\sum_{q<j}CA^{j-1-q}(B\bar u_{t+q}+e_{t+q})$,
and splitting $G_j\bar u$ in \eqref{eq:zhat} into the parts due to
$\uini$ and to $\bar u_f$ gives
$\yc{j}=CA^j\hat z_t+\sum_{q<j}CA^{j-1-q}B\bar u_{t+q}$; their
difference is \eqref{eq:struct}. Iterating \eqref{eq:resid} over the
window gives
$\bar z_t=A^{\Tini}\bar z_{t-\Tini}+\mathcal{R}\uini
+\sum_{s}A^{\Tini-1-s}e_{t-\Tini+s}$, and
$\hat z-\bar z_{t-\Tini}=\mathcal{O}^{\dagger}\dini$ from the proof of
Lemma~\ref{lem:error} gives \eqref{eq:est}.
\end{proof}

The first term of \eqref{eq:struct}, the error of estimating the
current lifted state from the window propagated $j$ steps, is fixed by
residuals that have already occurred, all but $e_{t-1}$ of which shaped
the window. The second collects the residuals along the predicted
trajectory, which have not occurred yet. Both are small in quiet
operation and typically large after a disturbance, which the window
shows. The next section therefore uses a window-dependent margin
calibrated on closed-loop data instead of a residual bound.

\section{Main Results: \texorpdfstring{\SC}{SC-DeePC} via Conformal Robustness}
\label{sec:bound}

Section~\ref{sec:pred} showed that the error $\err_j(t)$ of the
corrected prediction depends on neither $\alpha$ nor $\sigma_p$, so the
optimizer can change it only through the input and a bound on it can be
fixed before the problem is solved, and that it is driven by residuals
that the measured window partly reveals. Instead of bounding it for every
operating condition, this section bounds it with a prescribed
probability over the conditions the controller meets and calibrates the
bound on closed-loop data. This is the concept of \emph{conformal
robustness} borrowed from \cite{hsu2025conformal}, carried to DeePC
through the corrected prediction and its error identity
\eqref{eq:error}.
Section~\ref{sec:shape} learns a window-dependent shape for the margin.
Section~\ref{sec:cp} sets its size by conformal calibration on whole
closed-loop scenarios, which gives a guarantee that holds at every
instant of a new scenario simultaneously (Theorem~\ref{thm:cp}).
Section~\ref{sec:ctrl} tightens the output constraint of DeePC by the
calibrated margin, which yields \SC\ and its certificate
(Proposition~\ref{prop:cert}).

\subsection{A window-dependent margin}
\label{sec:shape}

The controller certifies the steps $j=1,\dots,H$, $H\le N-1$. By
Corollary~\ref{cor:struct}, the margin for step $j$ must cover the
propagated estimation error and the residuals along the horizon. Both
are typically large when the window shows a disturbance, which a
window-dependent margin exploits. The margin is written as a shape times
a size,
\begin{equation}
  \beta_{j,i}(t)=\gamma\cdot\sigma_\theta(w_t)_{j,i},
  \label{eq:beta}
\end{equation}
where $w_t$ is the window $\yini$ with the angles referred to their mean,
since the common angle mode is inert and drifts. The \emph{shape}
$\sigma_\theta$ maps the window to one positive number per certified
step and machine and describes how the error varies with the window.
The \emph{size} $\gamma$ is one scalar that makes the margin valid.
Separating the two lets any learning method propose the shape while the
guarantee rests on $\gamma$ alone; a poor network gives a
wider margin, not an invalid one.

The shape is a recurrent network, a gated recurrent unit
\cite{chung2014gru} followed by a small feed-forward head, because the
window is a short multivariate time series whose latest samples matter
most. The network is trained with the pinball loss to predict a high
quantile of $\log|\err_{j,i}|$, and $\sigma_\theta$ is the exponential
of its output, so that $\sigma_\theta(w_t)_{j,i}$ is positive and
approximates a conditional quantile of $|\err_{j,i}|$ given the window;
the logarithm lets it follow errors spanning orders of magnitude.

\subsection{Conformal calibration for conformal robustness}
\label{sec:cp}

A \emph{scenario} is one simulated closed-loop run, consisting of an
operating point, an event and the response under the controller. After
the run, the label at $(t,j,i)$ is $|\err_{j,i}(t)|$ computed from the
stored prediction and the outputs that actually occurred,
which is the error of Lemma~\ref{lem:error} plus, for $j\ge2$, the effect
of inputs re-planned after $t$. For each scenario $r$ compute the single
score
\begin{equation}
  R_r:=\max_{t}\ \max_{1\le j\le H}\ \max_{i}\
  \frac{|\err_{j,i}(t)|}{\sigma_\theta(w_t)_{j,i}},
  \label{eq:score}
\end{equation}
the maximum over all instants at which the controller solved, over the
certified steps and over the machines; $R_r$ is the smallest $\gamma$
that would have covered that entire scenario. Given $n$ calibration
scenarios, let $\hat\gamma$ be the $\lceil(n+1)(1-\epsilon)\rceil$-th
smallest of $R_1,\dots,R_n$.

The score is taken per scenario, not per instant, for two reasons. The
controller decides online at which instants to certify, so a guarantee
for a fixed instant does not suffice; it could fail at exactly the
instants the controller selects. The maximum in \eqref{eq:score}
makes the guarantee hold at every instant simultaneously, including the
certified ones. As in trajectory-level conformal prediction
\cite{hsu2025conformal}, it also avoids treating overlapping windows of
one run as independent samples.

\begin{assumption}\label{as:exch}
The calibration scenarios and the new scenario are exchangeable draws
from $\mathcal{D}$ and run under the same controller, and
$\sigma_\theta$ was trained on data disjoint from the calibration
scenarios.
\end{assumption}

The following is the split conformal guarantee
\cite{vovk2005algorithmic,angelopoulos2023conformal} applied at the
level of whole trajectories, as in \cite{hsu2025conformal}; what is
specific here is the score \eqref{eq:score}.
\begin{theorem}[{[\citen{vovk2005algorithmic}]}]\label{thm:cp}
Let Assumption~\ref{as:exch} hold, let $n\ge1/\epsilon-1$, and set
$\gamma=\hat\gamma$ in \eqref{eq:beta}. For a new scenario
from $\mathcal{D}$,
\begin{equation}
  \Pr\Bigl\{\,|\err_{j,i}(t)|\le\beta_{j,i}(t)\ \ \forall\, t,\
  \ 1\le j\le H,\ i\,\Bigr\}\ \ge\ 1-\epsilon ,
  \label{eq:guarantee}
\end{equation}
where the probability is taken jointly over the $n$ calibration
scenarios and the new scenario.
\end{theorem}
\begin{proof}
The event in \eqref{eq:guarantee} is $R_{n+1}\le\hat\gamma$ for
the exchangeable scores $R_1,\dots,R_{n+1}$; see
\cite{vovk2005algorithmic}.
\end{proof}

Theorem~\ref{thm:cp} is the conformal half of conformal robustness.
With probability at least $1-\epsilon$, the calibrated margin covers the
error of the corrected prediction at every instant and certified step of
a new scenario. The robust half is the controller of
Section~\ref{sec:ctrl}. At a certified instant it keeps the corrected
prediction inside the band shrunk by the margin, so that any error
within the margin keeps the next output inside the band
(Proposition~\ref{prop:cert}).

\begin{remark}\label{rem:shift}
The labels are collected with \SC\ running with $\beta=0$, whereas the
deployed loop runs with the margin, which changes the input wherever the
tightening is active and is scaled by $\hat\gamma$, so the deployed loop
also depends on the calibration set. Assumption~\ref{as:exch} therefore
holds for the deployed loop only approximately, and the coverage
measured in Section~\ref{sec:results} is the empirical check. Conformal
guarantees under such departures from exchangeability are studied in
\cite{barber2023beyond}.
\end{remark}

\subsection{Safety-certified DeePC}
\label{sec:ctrl}

This subsection supplies the robust half of conformal robustness. \SC\
is regularized DeePC with the corrected prediction in the cost and in a
tightened output constraint,
\begin{subequations}\label{eq:qp}
\begin{align}
  \min_{\alpha,\sigma_p,\bar u_f,s}\ &
  \sum_{j=0}^{N-1}\|S\yc{j}\|_Q^2
  +R_u\|\bar u_f\|^2+R_\Delta\|\Delta\bar u_f\|^2\nonumber\\
  &\quad+R_{\mathrm{seq}}\|\bar u_f-\bar u_f^{\,\mathrm{prev}}\|^2
  +\lambda_\alpha\|\alpha\|^2\nonumber\\
  &\quad+\lambda_\sigma\|\sigma_p\|^2
  +w_s\mathbf{1}\trans s \label{eq:qp-cost}\\
  \text{s.t. }\ & \eqref{eq:consist},\quad
  |\bar u_k|\le u_{\max},\quad |\Delta\bar u_k|\le\Delta u_{\max},
  \label{eq:qp-in}\\
  & |S\,\yc{j}|\le\band\mathbf{1}-b_j(t)+s_j\mathbf{1},\quad
  s_j\ge0, \label{eq:tight}
\end{align}
\end{subequations}
for $j=0,\dots,N-1$, where $\Delta\bar u_f$ collects the input
increments (the first relative to the last applied input),
$\bar u_f^{\,\mathrm{prev}}$ is the previous solution shifted by one
step, and the tightening is $b_{j,i}(t)=\beta_{j,i}(t)$ for
$j=1,\dots,H$ whenever $\beta_{j,i}(t)<\band$, and $b_{j,i}(t)=0$
otherwise. A margin wider than the band cannot be met by any input, so
it is not imposed. The DeePC baseline uses the same cost with
$E_jY_f\alpha$ in place of $\yc{j}$, the same regularization, input
limits and slack penalty, and the soft band
$|SE_jY_f\alpha|\le\band\mathbf{1}+s_j\mathbf{1}$ in place of
\eqref{eq:tight}. Since $\yc{j}$ is affine in the decision
variables, \eqref{eq:qp} is a convex quadratic program in which $b(t)$
is a parameter. The slack $s$ keeps the problem feasible
when no admissible input can meet the band, and it is heavily
penalized.

\textbf{Certified instant.} A sampling instant $t$ is \emph{certified}
if the full margin was imposed on every certified step, i.e.,
$\beta_{j,i}(t)<\band$ for all $j\le H$ and all $i$, and \eqref{eq:qp}
was solved with $s_j=0$ for $j=1,\dots,H$. A too-wide margin or a
positive slack therefore yields an uncertified instant, never a false
claim; hence the certificate is conditional.

\begin{proposition}[]\label{prop:cert}
Let instant $t$ be certified and let the margin hold at $t$, i.e.,
$|\err_{j,i}(t)|\le\beta_{j,i}(t)$ for $j=1,\dots,H$ and all $i$. Then
$|(S\bar y_{t+1})_i|\le\band$ for all $i$. If, in addition, the planned
inputs $\bar u_{t+1},\dots,\bar u_{t+H-1}$ are applied, then
$|(S\bar y_{t+j})_i|\le\band$ for $j=1,\dots,H$.
\end{proposition}
\begin{proof}
For $j\le H$ the tightened constraint holds with zero slack, so
$|(S\yc{j})_i|\le\band-\beta_{j,i}(t)$. With the planned inputs
applied, prediction and output use the same inputs, so by
\eqref{eq:rho}
$|(S\bar y_{t+j})_i|\le|(S\yc{j})_i|+|\err_{j,i}(t)|\le\band$. The output
$\bar y_{t+1}$ depends only on inputs up to $\bar u_t$, always
applied, so the first claim needs no assumption on later inputs.
\end{proof}

Under receding horizon only $\bar u_t$ is applied. Combining the first
claim of Proposition~\ref{prop:cert} with Theorem~\ref{thm:cp}, if
Assumption~\ref{as:exch} holds for the deployed loop, then with
probability at least $1-\epsilon$, jointly over the calibration
scenarios and a new scenario, $\bar y_{t+1}$ is inside the band at
every certified instant $t$ of the new scenario, with the caveat of
Remark~\ref{rem:shift}. Nothing is claimed at uncertified instants or
about recursive feasibility.

\begin{remark}[Relation to Koopman MPC]\label{rem:why}
By \eqref{eq:zhat}, \SC\ is a Koopman MPC \cite{kordamezic2018} whose
lifted state is estimated from the window, written in DeePC variables;
its terms in $\alpha$ and $\sigma_p$ add a convex penalty on
$\bar u_f$. The DeePC form is kept because removing the correction and
the margin returns regularized DeePC. The correction is needed because,
with $D_p:=Y_p-Y^0_p$ and $D_f:=Y_f-Y^0_f$, \eqref{eq:corrected} and
\eqref{eq:consist} give
$E_jY_f\alpha=\yc{j}+K_j\sigma_p+(E_jD_f-K_jD_p)\alpha$; a tightened
DeePC band could thus be met by moving $\alpha$ and $\sigma_p$ instead
of the input.
\end{remark}

\section{Simulation Setup}
\label{sec:setup}

\subsection{\texorpdfstring{Observables and nominal Koopman model}{Observables and nominal Koopman model}}
\label{sec:obs}

$\Phi$ collects each machine's rotor angle, speed and electrical
power, relative to the operating point,
\begin{equation}
  \Phi=\col\bigl(\delta_i-\delta_{0,i},\ \omega_i-\omega_s,\
  P_{e,i}-P_{e,i}^{0}\bigr),\qquad n_z=9,
  \label{eq:phi}
\end{equation}
so that \eqref{eq:swing1}--\eqref{eq:swing2} are exactly linear in
$\Phi$ and $u$. Trigonometric observables of the angle differences
\cite{susuki2011nonlinear}, a learned dictionary, thin-plate-spline
features \cite{kordamezic2018} and a Lie-derivative closure
\cite{netto2021observables} were fitted on the same samples. They
predict about equally well, but the largest entry of $K_j$, the noise
gain of the corrected prediction, is $3.5$ for \eqref{eq:phi} against
$67$ or more for the other sets. The one-step pairs ($12\,000$) are drawn on the region
$|\delta_i-\delta_{0,i}-c|\le0.2$~rad about their mean $c$,
$|c|\le0.3$~rad, $|\omega_i-\omega_s|\le\band$,
$|u_i|\le u_{\max}$ of the intact network at nominal load, and
$\bar e$ comes from an independent draw on the same
region. On a live grid, which cannot be placed at chosen states, the
pairs would come from recorded trajectories, and the lifted states
that start the columns of $Y^0_p,Y^0_f$ from recorded values of
\eqref{eq:phi}, whose electrical powers are routinely metered at the
generator terminals. The Hankel
record is one sequence of $304$ samples of the intact network at
nominal load under white Gaussian input noise of standard deviation
$0.05$~p.u., giving $265$ columns of depth $\Tini+N=40$; $\mathcal{O}$
has full column rank.

\subsection{Scenarios}

Five kinds of events are simulated. Three-phase \emph{faults} are
applied at buses 4--9, with shunt admittances of $1$, $10$ and
$100$~p.u., and cleared after $80$ or $250$~ms, which gives $108$ runs.
Fifteen runs have a permanent \emph{line outage}. Thirty runs have a
$10$~p.u., $80$~ms fault at one end of a line, which then trips and
\emph{recloses} after $0.4$ or $0.8$~s. Six runs have a \emph{load step} of $\pm15\,\%$, and six runs
are \emph{quiet}. Faults, outages, reclosures and load steps are
simulated at three pre-fault loadings, $85$, $92.5$ and $100\,\%$, with
the equilibrium re-solved for each loading and machine~1 acting as the
slack machine. The step-up transformers (lines $1$--$4$, $2$--$7$ and
$3$--$9$) are never tripped, five of the six remaining lines are
tripped in the outage and reclosure runs, and line $5$--$7$ is
\emph{held out}. Its trip appears in no training, calibration or test
scenario and is used only for the test of Section~\ref{sec:results},
outside the guarantee. Each run
lasts $3.5$~s, and the event starts at a random time between $0.6$ and
$1.4$~s. The $165$ scenarios are split at random into $66$ for
training, $16$ for validation (used only to stop training), $41$ for
calibration and $42$ for testing. The random split makes the
calibration and test scenarios exchangeable, which is what the proof of
Theorem~\ref{thm:cp} uses. It also leaves $28$ faults, $9$ reclosures,
$3$ load steps, one outage and one quiet run in the test set, so the
test results are dominated by faults and reclosures.

\subsection{Methods, metrics and parameters}

Four controllers are compared. The baseline is DeePC. The other three
are \SC\ with no margin ($\beta=0$), which isolates the corrected
prediction; with a \emph{constant} margin that ignores the window,
calibrated in the same way with $\sigma_\theta$ replaced by the $99$th
percentile, per step and machine, of the training labels; and with the
learned margin. All four share the same cost weights, input limits and
slack penalty.

\emph{Coverage} is the fraction of test scenarios in which the margin
held at every solved instant, certified step and machine, i.e., the
event in \eqref{eq:guarantee}, with target $1-\epsilon$. The
\emph{certified} fraction is the share of solved instants that are
certified, an instant being \emph{solved} if its window is full. \emph{Violations} are counted in machine-samples outside
the band, and a violation is \emph{claimed} if it occurs at $t+1$ after
a certified instant $t$; by Proposition~\ref{prop:cert} this happens
only where the margin fails.

\emph{Parameters}: sampling period $19.44$~ms, $\band=0.5$~Hz,
$u_{\max}=0.4$~p.u., $\Delta u_{\max}=0.05$~p.u., $\Tini=15$, $N=25$,
$H=4$ ($78$~ms), $\epsilon=0.05$, $Q=1/\band^2$ on the frequency rows,
$R_u=0.1$, $R_\Delta=5$, $R_{\mathrm{seq}}=50$, $\lambda_\alpha=1$,
$\lambda_\sigma=10^3$, $w_s=10^6$; no measurement noise.
$\sigma_\theta$ has one recurrent layer of width $24$ and a head of one
hidden layer of $64$ ReLU units, the pinball-loss level is $0.99$, and
$\hat\gamma$ is the $40$th smallest of the $41$ calibration scores.

\section{Results}
\label{sec:results}

\begin{table}[t]
\caption{The $42$ test scenarios from $\mathcal{D}$. Margin: median
over scenarios and solved instants of the largest entry of $\beta$.}
\label{tab:main}
\centering
\captabgap
\setlength{\tabcolsep}{2.6pt}
\begin{tabular}{lrrrrrr}
\toprule
method & margin & coverage & certified & violations & claimed & peak \\
       & (Hz)   & (\%)     & (\%)      &            &         & (Hz) \\
\midrule
DeePC & --- & --- & --- & 545 & --- & 1.84 \\
\SC, $\beta=0$ & 0 & --- & --- & 313 & --- & 1.51 \\
\SC, constant & 1.42 & 100.0 & 0.0 & 378 & 0 & 1.51 \\
\SC, learned & 0.45 & 90.5 & 47.1 & 356 & 0 & 1.51 \\
\bottomrule
\end{tabular}
\end{table}

\begin{table}[t]
\caption{Held-out line 5--7; final: deviation at $30$~s.}
\label{tab:stress}
\centering
\captabgap
\setlength{\tabcolsep}{3pt}
\begin{tabular}{lrrrr}
\toprule
method & certified (\%) & violations & peak (Hz) & final (Hz) \\
\midrule
DeePC & --- & 4325 & 10.5 & runs away \\
\SC & 87.2 & 6 & 0.56 & $-0.31$ \\
\bottomrule
\end{tabular}
\end{table}

\begin{figure*}[!t]
\centering
\subfloat[Machine 1\label{fig:traj1}]{\includegraphics[width=0.32\textwidth]{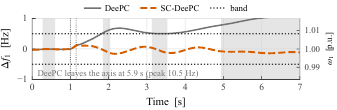}}\hfill
\subfloat[Machine 2\label{fig:traj2}]{\includegraphics[width=0.32\textwidth]{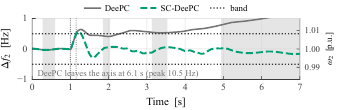}}\hfill
\subfloat[Machine 3\label{fig:traj3}]{\includegraphics[width=0.32\textwidth]{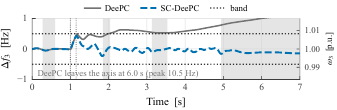}}\\[2pt]
\subfloat[Input of machine 2\label{fig:input}]{\includegraphics[width=0.32\textwidth]{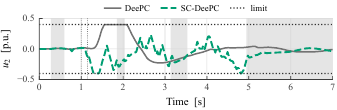}}\hfill
\subfloat[Machine 2, line 5--7 restored\label{fig:intact}]{\includegraphics[width=0.32\textwidth]{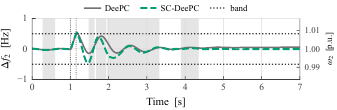}}\hfill
\subfloat[Next-step margin and error\label{fig:margin}]{\includegraphics[width=0.32\textwidth]{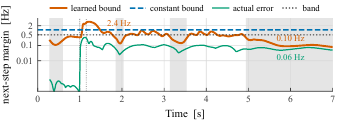}}
\caption{Test of Table~\ref{tab:stress}; certified instants shaded.
(a)--(c) Frequency deviation of each machine: DeePC solid grey, \SC\
dashed, band dotted. (d) Control input of machine 2; dotted lines are
the $\pm u_{\max}$ limits. (e) Machine 2 for the same fault with line
5--7 restored at $1.15$~s. (f) Learned margin $\max_i\beta_{1,i}(t)$,
constant margin and actual next-step error $\max_i|\err_{1,i}(t)|$, in
Hz on a log scale.}
\label{fig:heldout}
\end{figure*}

We first compare the bounds of Proposition~\ref{prop:wc} with the
learned margin, then report the test set, the held-out line and all
lines. The largest entry
of $\beta^{\mathrm{sep}}$ in \eqref{eq:det} exceeds
the band at every step, one step ahead by a factor of about three. That
of the exact worst case $\beta^{\mathrm{ex}}$ is four to six times
smaller on the certified steps, $0.24$~Hz one step and $0.38$~Hz four
steps ahead, and fits inside the band there, but it is guaranteed only
while the residuals stay in the box. One step ahead, where the realized
error is that of Lemma~\ref{lem:error}, it covers $15$ of the $42$ test
scenarios on their $\beta=0$ runs, failing in $18$ of the $28$ faults
and all $9$ reclosures with errors up to $3.1$ times the bound, whereas
the learned margin covers $39$ on the same runs and step. The identity \eqref{eq:error} holds to
machine precision on closed-loop data through a fault.

\subsection{Test set}
The learned margin covers $38$ of the $42$ test scenarios
(Table~\ref{tab:main}). For i.i.d.\ continuous scores the coverage
given the calibration scenarios is $\mathrm{Beta}(40,2)$
\cite{vovk2005algorithmic,angelopoulos2023conformal}, so the number of
uncovered scenarios among $42$ has mean $2.0$ and is four or more with
probability $0.19$; four misses are consistent with the guarantee,
whose assumption holds only approximately
(Remark~\ref{rem:shift}). In the four uncovered
scenarios the margin is exceeded only at uncertified instants, and no
certified instant is followed by a sample outside the band.

The learned margin certifies $47\,\%$ of the solved instants, and its
median size in Table~\ref{tab:main} is $0.45$~Hz. The constant margin
has largest entry $1.42$~Hz and is
narrower than the band on machine~1 only, so it never certifies an
instant although it covers every scenario.  

Compared with DeePC, \SC\ with the learned margin reduces the band
violations by $35\,\%$ and the peak deviation from $1.84$ to $1.51$~Hz.
All of this comes from the corrected prediction alone ($\beta=0$), which
reduces the violations by $43\,\%$. The remaining violations occur
during fault and reclosure transients, which
start without warning and which the bounded input does not stop; the
other scenarios stay inside the band under every controller. The peak
of $1.51$~Hz, common to the three \SC\ variants, comes from a
$100$~p.u.\ fault at bus~7 and occurs as the fault is cleared after
$250$~ms. A margin is imposed wherever it fits inside the band, which
during a transient is often on some machines only; the constant margin
fits on machine~1 alone. Such a tightened constraint often cannot be met
during the transient, and since the slack $s_j$ in \eqref{eq:tight} is
shared by the machines, the slack it requires also relaxes the band on
the machines that swing. This slightly increases the violations relative
to $\beta=0$, but none follows a certified instant.

\subsection{Held-out line}
Table~\ref{tab:stress} and Fig.~\ref{fig:heldout}(a)--(d) show
the fault that trips the held-out line, whose post-trip network appears
in no training data: a $22$~p.u.\ fault at bus~5 at $1.0$~s,
cleared at $1.15$~s by tripping line 5--7, which stays out for the rest
of the $30$~s run. DeePC re-enters the band after the clearing
transient, drifts out again within a second and saturates its actuators
(Fig.~\ref{fig:heldout}(d)), and the common frequency of the machines
runs away. \SC\ leaves the band only during the clearing transient, with
six violations on machine~2 between $1.17$ and $1.26$~s, none following
a certified instant. It then settles about $0.31$~Hz below nominal and
certifies most instants.

In this run the learned margin (Fig.~\ref{fig:heldout}(f)) widens to about
$2.4$~Hz during the fault and the first half second after the trip, and
stays near the band while the post-trip oscillation decays. From $5$~s
every instant is certified. The margin falls from about $0.24$~Hz at
$5$~s to about $0.1$~Hz at $5.7$~s and then follows the actual next-step
error of about $0.06$~Hz within a factor of about $1.6$. With the line
restored
instead (Fig.~\ref{fig:heldout}(e)), \SC\ has the smaller peak on every
machine and returns to zero. One solve takes about a second against a
$19.44$~ms period, so real-time operation is left for future work.

\subsection{All switchable lines}

Table~\ref{tab:lines} repeats this test for the six lines that are not
step-up transformers, with the fault at the first-named bus and
runs of $20$~s; in $\mathcal{D}$ the other five trip only without
a fault or with reclosure, so Theorem~\ref{thm:cp} does not apply. DeePC
leaves the band after every trip and is farthest out at the end of each
run, while \SC\ ends within $0.37$~Hz of the nominal frequency and has
no violation after $5$~s; its peaks of $0.50$~Hz are just above the
band. After
the trip of line 8--9, however, the four certified
instants from $1.575$ to $1.633$~s are each followed by a violation on
machine~3, of at most $0.64$~Hz. There the largest one-step margin over
machines, $0.09$ to $0.11$~Hz, is below the largest one-step error,
$0.12$ to $0.25$~Hz, so the certificate is wrong at these instants.
They are the only failed certificates in this paper, on a run
outside $\mathcal{D}$.

\begin{table}[t]
\caption{Trip of each switchable line.}
\label{tab:lines}
\centering
\captabgap
\setlength{\tabcolsep}{4pt}
\begin{tabular}{lrrrr}
\toprule
& \multicolumn{2}{c}{DeePC} & \multicolumn{2}{c}{\SC}\\
\cmidrule(lr){2-3}\cmidrule(lr){4-5}
line & peak (Hz) & violations & peak (Hz) & violations\\
\midrule
4--5 & 3.88 & 1979 & 0.50 & 1\\
4--6 & 1.98 & 1497 & 0.50 & 1\\
5--7, held out & 7.25 & 2783 & 0.56 & 6\\
6--9 & 5.97 & 2238 & 0.49 & 0\\
7--8 & 1.74 & 1542 & 0.84 & 28\\
8--9 & 2.99 & 1630 & 0.66 & 32\\
\bottomrule
\end{tabular}
\end{table}

\section{Conclusion}

This paper introduced \SC, a DeePC controller for nonlinear plants
with conditional run-time safety certificates, built on an exact error
identity and, borrowing conformal robustness, a window-dependent margin
calibrated by conformal prediction. On the IEEE nine-bus benchmark, the
corrected prediction alone reduces the violations and the peak
deviation of DeePC, and the learned margin makes certification possible
where a window-blind margin certifies no instant. Certificates are
withheld while the margin does not fit inside the band, as in fault
transients, and none issued on the test scenarios or after the
held-out outage is followed by a violation. After that outage, \SC\
stays in the band beyond the clearing transient and certifies every
instant once the plant settles, while DeePC runs away. Future work will
address recursive feasibility \cite{berberich2021data} and extend
conformal robustness \cite{hsu2025conformal} to certificates beyond the
next sample.

\bibliographystyle{IEEEtran}
\bibliography{refs}

\end{document}